# Interstellar travels aboard radiation-powered rockets

André Füzfa[*]
*Namur Institute for Complex Systems (naXys), University of Namur,
Rue de Bruxelles 61, B-5000 Namur, Belgium*

We model accelerated trips at high velocity aboard light sails (beam-powered propulsion in general) and radiation rockets (thrust by anisotropic emission of radiation) in terms of Kinnersley's solution of general relativity and its associated geodesics. The analysis of radiation rockets relativistic kinematics shows that the true problem of interstellar travel is not really the amount of propellant nor the duration of the trip but rather its tremendous energy cost. Indeed, a flyby of Proxima Centauri with an ultralight gram-scale laser sail would require the energy produced by a 1 GW power plant during about one day, while more than 15 times the current world energy production would be required for sending a 100 tons radiation rocket to the nearest star system. The deformation of the local celestial sphere aboard radiation rockets is obtained through the null geodesics of Kinnersley's spacetime in the Hamiltonian formulation. It is shown how relativistic aberration and the Doppler effect for the accelerated traveler differ from their description in special relativity for motion at constant velocity. We also show how our results could interestingly be extended to extremely luminous events like the large amount of gravitational waves emitted by binary black hole mergers.

DOI: 

## I. INTRODUCTION

We would like to start this article by applying to physics Albert Camus's words from his essay [1]. There is but one truly serious *physical* problem and that is *interstellar travel*. All the rest—*whether or not spacetime has four dimensions, whether gravity is emergent or can eventually be quantized*—comes afterwards[1]. It is not that such questions are not important nor fascinating—and actually they could quite likely be related to the above-mentioned crucial problem—but interstellar travel is the most *appealing* physical question for a species of explorers such as ours. Interstellar travel, although not theoretically impossible, is widely considered as practically unreachable. This topic has also been often badly hijacked by science-fiction and pseudo-scientific discussion when not polluted by questionable works.

A simplistic view of the problem of interstellar travel is as follows. On one hand, the distance to the closest star system—Alpha Centauri—is roughly 4 orders of magnitude larger than the approximately 4.5 billion km toward planet Neptune. On the other hand, the highest velocity a man-made object has ever reached so far [2] is only of the order 10 km/s. Consequently, it would roughly take more than 100 millennia to get there with the same technology. To cross interstellar distances that are of order of light years (1 light year being approximately $9.5 \times 10^{12}$ km) within human timescales, one must reach relativistic velocities, i.e., comparable to the speed of light. However, from the famous Tsiolkovsky rocket equation,

$$\Delta v = v_e \log\left(\frac{m_i}{m_f}\right) \quad (1)$$

(with $\Delta v$ being the variation of rocket velocity during ejection of gas with exhaust velocity $v_e$ and $m_{i,f}$ being the initial and final total mass), any space vehicle accelerated by ejecting some mass that eventually reaches a final velocity $\Delta v$ of $10,000$ km/$s$ of the speed of light $c$ with an exhaust velocity $v_e$ of 1 km/s, typical of a chemical rocket engine, would require an initial mass of propellant $m_i$ about $10^{4000}$ larger than the payload $m_f$. To reduce the initial mass $m_i$ to 100 times the mass of the payload $m_f$ with the final velocity $\Delta v \sim 0.1 \times c$ requires exhausting mass at a speed of approximately $6 \times 10^3$ km/s, about 4 orders of magnitude higher than with current conventional rocketry. With this quick reasoning, where Eq. (1) is based on Newtonian dynamics, one would conclude that interstellar travel would require either 10 000 times longer trips or 10 000 times faster propulsion. There have been many suggestions (sometimes based on questionable grounds) and engineering preliminary studies for relativistic spaceflight, and we refer the reader to Ref. [3] and references therein for an overview. Recently, there has been some renewed interest on relativistic spaceflight using photonic propulsion, along with the Breakthrough Starshot project [4] aiming at sending nanocraft toward Proxima Centauri by shooting high-power laser pulses on a light sail to which the probe will be attached.

---

[*]andre.fuzfa@unamur.be
[1]There is but one truly serious philosophical problem and that is suicide. (…) All the rest—whether or not the world has three dimensions, whether the mind has nine or twelve categories—comes afterwards.

According to us, such a quick analysis as above somehow hides the most important problem—the energy cost—really preventing interstellar travel from becoming a practical reality. Both issues of the duration of the trip and the propellant can be fixed, at least in principle. First, time dilation in an accelerated relativistic motion reduces the duration of any interstellar trip [5], though this requires delivering an enormous total amount of energy to maintain some acceleration all throughout the trip. Second, space propulsion can be achieved without using any (massive) propellant (and without any violation of Newton's third law): light and gravitational waves are carrying momenta, and their emission exerts some reaction recoil of the source. General relativity allows modeling the accelerated motion of a particle emitting or absorbing radiation anisotropically through a special class of exact solutions called the *photon rocket spacetimes*.

In 1969, Kinnersley published in Ref. [6] a generalization of Vaidya's metric, which is itself a generalization of the Schwarzschild exterior metric. Vaidya's solution [7] represents the geometry of spacetime around a pointlike mass in inertial motion that is either absorbing or emitting radiation, through a null dust solution of Einstein's equation:

$$R^{\mu\nu} = \kappa \Phi k^\mu k^\nu \qquad (2)$$

(with $R^{\mu\nu}$ being the Ricci tensor, $\Phi$ being a scalar field describing the radiation flux, and $k^\mu$ being a null vector field). Kinnersley's solution [6] describes spacetime around such a radiating mass undergoing arbitrary acceleration due to anisotropic emission and contains four arbitrary functions of time: the mass of the point particle $m$ and the three independent components of the 4-acceleration $a^\mu$ of the particle's worldline. It further appeared that Kinnersley's solution was only a special case of a more general class of exact solutions of general relativity, the Robinson-Trautman spacetimes (cf. Ref. [8] for a review). The term "photon rockets" for such solutions was later coined by Bonnor in Ref. [9], although it would be preferable to use the term "radiation rocket" to avoid any confusion with quantum physics vocabulary [10]. The literature on radiation rockets focused mainly on exploring this class of exact solutions or extending Kinnersley's solution (see Refs. [8,11,12] and references therein), notably through solutions for gravitational wave–powered rockets [13], (anti-)de Sitter background [12], and photon rockets in arbitrary dimensions [14]. Another interesting point about Kinnersley's solution is the fact that there is no gravitational radiation emission (at least at the linear level) associated to the acceleration of the point mass by electromagnetic radiation, as explained in Ref. [15].

In Ref. [14], the first explicit solutions of photon rockets were given, as was a detailed presentation of several useful background Minkowski coordinates for describing the general motion of the particle rocket. The solutions given in Ref. [14] for straight flight includes the case of hyperbolic motion (*constant acceleration*) and another analytical solution, both in the case of an emitting rocket, for which the rocket mass decreases. The case of light sails (either solar or laser-pushed sails, i.e., a photon rocket which absorbs and reflects radiation) is not considered while, as we show here, can be described by the same tool. Geodesic motion of matter test particles is also briefly discussed in Ref. [14] to confirm the absence of gravitational aberration investigated in Ref. [16].

In the present paper, we apply Kinnersley's solution to the modeling of relativistic motion propelled either by anisotropic emission or absorption of radiation. This encompasses many photonic propulsion proposals like blackbody rockets [17] (in which a nuclear source is used to heat some material to high temperature and its blackbody radiation is then appropriately collimated to produce thrust), antimatter rockets, or light sails [3]. Spacetime geometry in traveler's frame is obtained through the resolution of energy-momentum conservation equations, which results in decoupled rocket equations for the acceleration and the mass functions, while standard approach using Einstein equations deals with a single constraint mixing together the radiation source characteristics and the kinematical variables. This approach also allows working directly with usual quantities such as emissivity, the absorption coefficient, and the specific intensity of the radiation beam acting on the particle to derive the corresponding relativistic kinematics. We provide simple analytical solutions for the case of constant radiative power. We also rederive the relativistic rocket equation once obtained in Ref. [18] by a more simplistic analysis with basic special relativity. We then apply the radiation rocket kinematics to three detailed toy models of interstellar travel: (i) the Starshot project, an interstellar flyby of a gram-scale probe attached to a light sail that is pushed away by a ground-based laser, and (ii,iii) single and return trips with an emission radiative rocket of mass scale 100 t. We show how the unreachable energy cost of the latter strongly speaks in favor of the former. Then, from the associated spacetime geometry around the traveler, we investigate light geodesics through Hamiltonian formulation. Incoming and outgoing radial trajectories of photons are obtained and used to compute frequency shifts, for instance in telecommunications between the traveler, his home, and his destination. We then investigate the deviation of the angle of incidence of incoming photons—as perceived by the traveler as it moves. We show how local celestial sphere deformation evolves with time in a different way for either an emission or an absorption rocket and in a different way than in special relativity, establishing how this effect constitutes a crucial question for interstellar navigation which cannot be captured by special relativity alone.

The layout of this paper is as follows. In Sec. II, we set the basics of radiation rockets in general relativity before we introduce our procedure to establish explicit solutions given incoming (respectively outgoing) radiation corresponds to absorption (respectively emission) rocket(s). Relativistic photon rocket equations, analytical solutions, as well as numerical solutions for more realistic cases are derived. We then apply these solutions to three toy models of interstellar travel: flyby with a light sail and a single and return interstellar trips with an emission rocket. Section III is devoted to the study of geodesics in Kinnersley spacetime using Hamiltonian formulation. Relativistic aberration and the Doppler effect for the accelerated observer are obtained and compared to the case of an observer moving at constant speed as described by special relativity. Visualizations of interstellar panoramas for accelerated travelers resulting from relativistic aberration, the Doppler effect, and relativistic focusing are presented. We finally conclude in Sec. IV by emphasising key implications of our results for the problem of interstellar travel as well as introducing another possible application of the present results to the astrophysical problem of gravitational wave recoil by binary black hole mergers.

## II. SPACETIME GEOMETRY AROUND RADIATION ROCKETS

### A. Deriving the radiative rocket equations

In what follows, we build explicit solutions of the photon rocket spacetime. We first review some basics of so-called photon rockets in general relativity and refer the reader to Refs. [6,9,12,14,15] for alternative nice presentations. Our procedure to determine spacetime geometry around radiation rockets requires two frames and associated coordinates. The first observer $\mathcal{O}$ is located far from the radiation rocket so that the gravitational attraction of the (variable) rocket mass can be neglected and corresponds to an inertial frame with associated Cartesian coordinates $(X^\mu)_{\mu=0,\ldots,3} = (cT, X, Y, Z)$, where $c$ is the speed of light and is left to facilitate dimensional reasoning for the reader. The second observer $\mathcal{O}'$ is the traveler embarked on the radiation rocket, using comoving spherical coordinates $(x^\nu)_{\nu=0,\ldots,3} = (c \cdot \tau, r, \theta, \varphi)$, with $\tau$ being the proper time of the traveler.

We will assume throughout this paper that the trajectory of the rocket follows a straight path along the direction $Z$ in $\mathcal{O}$ coordinates so that the traveler's worldline $\mathcal{L}$ is given either by $\mathcal{L} \equiv (cT(\tau), 0, 0, Z(\tau))$ in $\mathcal{O}$'s coordinates or $\mathcal{L} \equiv (c \cdot \tau, r = 0)$ (and with any $\theta, \varphi$) in $\mathcal{O}'$ coordinates. The tangent vector field to the worldline $\mathcal{L}$ will be denoted by $\lambda^\mu = dX^\mu/(cd\tau)$ in $\mathcal{O}$ and $\lambda'^\mu = dx^\mu/(cd\tau)$ in $\mathcal{O}'$.

According to the equivalence principle, the accelerated traveler experiences a local gravitational field. The geometry of spacetime can be described in the comoving coordinates $(c \cdot u, r, \theta, \varphi)$ by the following metric,

$$ds^2 = c^2 \left(1 - 2\frac{M}{r} - 2\alpha r \cos(\theta) - \alpha^2 r^2 \sin^2(\theta)\right) du^2$$
$$\pm 2c du dr \mp 2\alpha r^2 \sin(\theta) c du d\theta$$
$$- r^2(d\theta^2 + \sin^2(\theta) d\varphi^2), \qquad (3)$$

where the signs in the second line above stand for the cases of retarded or advanced time coordinates $u$, respectively. One can indeed associate each point $X^\mu$ of Minkowski space to a unique retarded or advanced point $X^\mu_\mathcal{L}$ on the worldline $\mathcal{L}$ which is at the intersection of the past/future light cone of $X^\mu$ and the worldline $\mathcal{L}$. It should be noticed that in both limits of a vanishing mass $M \to 0$ or far away from the point rocket $r \to \infty$, the null coordinate $u$ takes the value of the proper time $\tau$ of observers aboard the rocket located at $r = 0$ in those comoving coordinates (we also refer the reader to Ref. [14] for a more detailed discussion). The retarded and advanced metrics will be used in Sec. III for computing incoming and outgoing geodesics toward/ from $X^\mu_\mathcal{L}$. The above metric is of course singular at the traveler's location $r = 0$, and the geometric quadratic invariants only depend on mass $M$ (see Ref. [9]). Therefore, when $M = 0$, one retrieves Minkowski spacetime but viewed by an accelerated observer.

Spacetime geometry (3) around the radiative rocket is totally specified by the two functions of time $M(u)$ and $\alpha(u)$. The first function $M(u) = 2Gm(u)/c^2$, where $G$ is Newton's constant, implements the gravitational effect of the inertial mass $m$ of the radiative rocket. The second function $\alpha(u)$ is related to the 4-acceleration of the radiative rocket in the following way [19]. Let $\dot{\lambda}^\mu = d\lambda^\mu/(cd\tau)$ be the 4-acceleration of the worldline $\mathcal{L}$, and since the unit tangent vector $\lambda^\mu$ is timelike ($\eta_{\mu\nu}\lambda^\mu\lambda^\nu = 1$), we have that $\dot{\lambda}^\mu \lambda_\mu = 0$ or in other words that $\dot{\lambda}^\mu$ is a spacelike vector. The function $\alpha(\tau)$ is then given by $\alpha^2 = -\dot{\lambda}^\mu \dot{\lambda}_\mu = -\dot{\lambda}'^\mu \dot{\lambda}'_\mu \geq 0$. As we shall see below, both mass $M$ and "acceleration" $\alpha$ functions will be linked together through the relativistic rocket equations.

One can move from traveler's comoving coordinates $\mathcal{O}'$ to inertial coordinates $\mathcal{O}$ through the transformation (see also Refs. [9,12,14] for the retarded case)

$$\begin{cases} T = T_\mathcal{L}(\tau) + r \cdot [\pm \cosh(\psi) + \cos(\theta) \cdot \sinh(\psi)] \\ X = r \cdot \sin(\theta) \cdot \cos(\varphi) \\ Y = r \cdot \sin(\theta) \cdot \sin(\varphi) \\ Z = Z_\mathcal{L}(\tau) + r \cdot [\pm \sinh(\psi) + \cos(\theta) \cdot \cosh(\psi)] \end{cases}, \qquad (4)$$

where the couple $(T_\mathcal{L}(\tau), Z_\mathcal{L}(\tau))$ is the functions specifying the worldline $\mathcal{L}$ of the traveler in inertial coordinates $\mathcal{O}$, $\psi = \psi(\tau)$ is the rapidity, and the case $+ (-)$ is for the retarded (advanced) metric, respectively. It must be kept in mind that this transformation is only valid for negligible radiative mass $m$, or at infinite distance from the rocket, and reduces to Eq. (3) in the limit $M \to 0$. Performing the transformation of metric (3) with $m = 0$ to Minkowski metric $\eta_{\mu\nu} = \text{diag}(+1, -1, -1, -1)$ yields

$$\lambda^T = \frac{dT_\mathcal{L}}{d\tau} = \cosh(\psi(\tau)) \quad (5)$$

$$\lambda^Z = \frac{dZ_\mathcal{L}}{d\tau} = c \cdot \sinh(\psi(\tau)) \quad (6)$$

$$\alpha(\tau) = -\dot{\psi}, \quad (7)$$

where a dot denotes a derivative with respect to $c \cdot \tau$. The tangent vector $\lambda^\mu = dX^\mu/(cd\tau) \equiv \dot{X}^\mu$ in inertial coordinates is normalized, $\eta_{\mu\nu}\lambda^\mu\lambda^\nu = 1$, and the (absolute value of the) norm of the 4-acceleration is given by $\dot{\lambda}_\mu\dot{\lambda}^\mu = \dot{\psi}^2$.

The usual approach to *photon rockets* is to pass by Einstein's equations, which reduces here to only one equation since the only nonvanishing component of the Ricci tensor for the metric (3) is written as

$$R^{11} = \frac{2}{r^2}\left(3\alpha M \cos(\theta) \mp \frac{dM}{cd\tau}\right), \quad (8)$$

and one can further verify that the scalar curvature $R$ identically vanishes. Equation (8) therefore puts a single constraint between the radiation flux function $\Phi$ of Eq. (2) and the kinematical functions of the mass $M$, its derivative $\dot{M} = dM/(cd\tau)$, and the 4-acceleration $\alpha$ of the radiation rocket. A decoupled set of radiation rocket equations would be more suitable for physical modeling since we would like to derive directly the worldline $\mathcal{L}$ and the associated spacetime geometry (3) from ingoing and outgoing radiation characteristics.

To achieve this, we follow Ref. [15] and consider total energy-momentum conservation in inertial frame $\mathcal{O}$, which is written as

$$\partial_\mu(T^{\mu\nu}_{(\text{m})} + T^{\mu\nu}_{(\text{rad})}) = 0. \quad (9)$$

On the one hand, we have that for a point particle of mass $m$

$$\partial_\mu T^{\mu\nu}_{(\text{m})} = c^2 \int d\tau \left[\frac{d}{d\tau}(m(\tau)\lambda^\mu)\delta^4(X^\mu - X^\mu_\mathcal{L}(\tau))\right],$$

where $X^\mu_\mathcal{L}(\tau)$ represents the location of the radiation rocket in spacetime coordinates of inertial frame $\mathcal{O}$. On the other hand, radiation's stress-energy conservation

$$\partial_\mu T^{\mu\nu}_{(\text{rad})} = -\mathcal{F}^\nu$$

defines the radiation reaction 4-force density acting on the rocket (see also Ref. [20]). The time component $\mathcal{F}^0$ gives $c^{-1}$ times the net rate per unit volume of radiative energy flowing into or escaping the particle, while the spatial components $\mathcal{F}^i$ give the thrust per unit volume that is imparted to the rocket. For a pointlike radiative distribution, one can write down

$$\mathcal{F}^\mu = \int d\tau f^\mu \delta^4(X^\mu - X^\mu_\mathcal{L}(\tau))$$

with $f^\mu$ the radiation reaction 4-force, which has units power. With these definitions, the conservation equation (9) yields the *relativistic radiation rocket equations* (10), (11)

$$\begin{cases} \dot{m}c^2 = \lambda_\mu f^\mu \\ mc^2 \dot{\lambda}^\mu = f^\mu - \lambda_\beta f^\beta \lambda^\mu, \end{cases} \quad (10)$$

where a dot now denotes a derivative with respect to $\tau$ and where we have used a contraction with $\lambda_\mu$ together with $\lambda_\mu\lambda^\mu = 1$. The radiation reaction 4-force $f^\mu$ is given by (see also Ref. [20])

$$f^\mu(\tau) = \int_0^\infty \oint \{[\mathcal{AI} - \mathcal{E}](\tau, \nu, \theta, \varphi)n^\mu(\theta, \varphi)\} \cdot d\nu \cdot d\Omega \quad (11)$$

with $\nu$ the radiation frequency, $d\Omega = \sin(\theta) \cdot d\theta \cdot d\varphi$ being the solid angle element, $\mathcal{A}$ being the absorption coefficient (with dimensions of an area in $m^2$), $\mathcal{I}$ being the specific intensity [with dimensions $W/(m^2 \, \text{Hz Ster})$], and $\mathcal{E}$ being the emission coefficient [with dimensions $W/(\text{Hz Ster})$] and $n^\mu = (1, \vec{n}(\theta, \varphi))$ is a dimensionless null 4-vector with $\vec{n}(\theta, \varphi)$ pointing outward the rocket, in the direction $(\theta, \varphi)$ of the unit 2-sphere:

$$\vec{n}(\theta, \varphi) = (\sin(\theta)\cos(\varphi), \sin(\theta)\sin(\varphi), \cos(\theta))^T.$$

In Eq. (11), the component $f^T$ gives the power that is either entering the rocket $f^T > 0$ (for a light sail or in other words an absorption rocket) or fleeing the rocket $f^T < 0$ (for an emission rocket), while the components $f^i$ are $c$ times the thrust imparted by the radiation flux to the rocket. The kinematic part of radiation rocket equations, Eqs. (10), actually describes general relativistic motion in the presence of some 4-force field $f^\mu$ (see also Ref. [5]). Therefore, specifying the radiation flux that is either emitted, reflected, or absorbed by the radiative rocket and the associated momentum gained by the rocket allows one to solve the kinematical equations (10), determining both the traveler's worldline $\mathcal{L}$ and the spacetime geometry through Eq. (3). We now focus on explicit solutions in the next paragraph.

## B. Straight accelerated motion of light sails, absorption, and emission radiation rockets

For a straight motion along the Z axis, one has $X = Y = \lambda^X = \lambda^Y = 0$ together with Eqs. (5) and (6) so that the relativistic rocket equations (10) reduce to

$$\begin{cases} \dot{m}c^2 = \cosh(\psi)f^T - \sinh(\psi)f^Z \\ mc^2\dot{\psi} = -\sinh(\psi)f^T + \cosh(\psi)f^Z. \end{cases} \quad (12)$$

The 3-velocity (i.e., velocity in the Newtonian sense) of the radiation rocket with respect to the inertial observer $\mathcal{O}$ is given by $V^Z = \frac{dZ}{dT} = c\frac{\lambda^Z}{\lambda^T} = c\tanh(\psi)$, while the Newtonian 3-acceleration with respect to the inertial observer $\mathcal{O}$ is given by $a = \frac{dV^Z}{dT} = c\dot{\psi}/\cosh^3(\psi)$ (a dotted quantity representing here the derivative of this quantity with respect to proper time $\tau$).

One can advantageously reformulate the system (12) in terms of dimensionless quantities by considering the characteristic scale for the proper time given by $\tau_c = m_0 c^2/|P|$ with $m_0$ being the inertial mass of the radiation rocket at start and $P$ being is the scale of the power driving the rocket and that is either entering ($P > 0$) or leaving the rocket ($P < 0$). Using $s = \tau/\tau_c$ as a dimensionless time variable, one can rewrite Eq. (12) as follows,

$$\begin{cases} \mathcal{M}' = \mathcal{P}(\tau).[\cosh(\psi) - \sinh(\psi)\mathcal{T}(\tau)] \\ \mathcal{M}\psi' = \mathcal{P}(\tau).[-\sinh(\psi) + \cosh(\psi)\mathcal{T}(\tau)], \end{cases} \quad (13)$$

where a prime denotes a derivative with respect to $s$, $\mathcal{M} = m/m_0$, $\mathcal{P}(\tau) = f^T/|P|$ is a dimensionless function describing the power either entering or leaving the rocket, and $\mathcal{T} = f^Z/f^T = T(\tau).c/f^T$ is a dimensionless function associated to the thrust (in Newtons) driving the radiation rocket.

Actually, there are two types of radiation rockets: absorption rockets, for which $\mathcal{P}(\tau) > 0$, and emission rockets, for which $\mathcal{P}(\tau) < 0$. Light or solar sails belong to the former, while a simple case of the latter consists of a system propelled by anisotropic radiative cooling (see also Ref. [17] for general introduction to the idea). If the thrust is directly proportional to the driving power, i.e., $\mathcal{T} = \pm 1$, the first of the relativistic rocket equation (13) reduces to

$$\mathcal{M}' = \mathcal{P}(\tau).\exp(\mp\psi).$$

Therefore, in the case of a purely absorbing rocket $\mathcal{P}(\tau) > 0$, the rocket mass $m$ is monotonically increasing, while in the case of a purely emitting rocket $\mathcal{P}(\tau) < 0$, the rocket mass $m$ is monotonically decreasing. Furthermore, it can be shown from Eqs. (13) with $\mathcal{T} = \pm 1$ that the 3-velocity of the radiation rocket with respect to the inertial observer $\mathcal{O}$, $V^Z = c\tanh(\psi)$ verifies

$$\Delta V^Z = c\left|\frac{m^2 - m_0^2}{m^2 + m_0^2}\right|, \quad (14)$$

which is nothing but a *relativistic generalization of the Tsiolkovsky equation* (1), as obtained for the first time by Ackeret in Ref. [18] from basic special relativity.

Let us now consider the simplistic case where the driving power is constant, $\mathcal{P}(\tau) = \pm 1$ and the thrust is directly proportional to this power, $\mathcal{T}(\tau) = \pm 1$. Under these assumptions, one can easily obtain the following analytical solution of the system (13):

$$\mathcal{M} = (1 + 2\mathcal{P}e^{-\mathcal{P}.\mathcal{T}\psi_0} \cdot (s - s_0))^{1/2} \quad (15)$$

$$\psi = \psi_0 + \mathcal{P}.\mathcal{T}\log\mathcal{M}, \quad (16)$$

where $s_0, \psi_0$ the initial dimensionless time and rapidity respectively, $\mathcal{P} = \pm 1$ (+1 for the absorption rocket and −1 for the emission one) and $\mathcal{T} = \pm 1$ gives the direction of acceleration (+1 acceleration toward +Z and −1 backward Z). One can easily check that Eqs. (15) and (16) of course satisfy Ackeret's equation (14). For an absorption rocket, the velocity will reach that of light $c$ asymptotically and with an infinitely large inertial mass, while an emission rocket ($\mathcal{P} = -1$) will reach the speed of light $c$ after a finite proper time $(s_{\text{rel}} - s_0) = e^{-\mathcal{T}\psi_0}/2$ where its inertial mass identically vanishes.

However, Eqs. (15) and (16) describe the radiation rocket powered by a constant source, which is too simplistic. First, since the intensity of a light beam decreases with the inverse of the distance to the source, the feeding power of a light sail will decrease as $\mathcal{P}(\tau) \sim Z^{-2}(\tau)$ unless the remote power source is increased accordingly, which is unpractical. Second, one could also consider an emission rocket propelled by directed blackbody radiation coming from the decay heat of some radioactive material, in which case the power source will decay as $\mathcal{P}(\tau) = \exp(-s/\mathcal{S})$ with decay time $\mathcal{S}$ (in units of characteristic proper time $\tau_c$). More realistic physical models of radiation rockets will therefore involve the time dependance of the rocket's driving power and thrust $\mathcal{P}, \mathcal{T}(\tau)$. Then, kinematics must be obtained in general through numerical integration of Eqs. (13).

We can now give two couples of simple models of light sails and emission rockets. For the light sails, let us assume the following: (i) The power function is given by $\mathcal{P}(\tau) = (1 + Z(\tau)/Z_c)^{-2}$ with $Z_c = c.\tau_c$ being a characteristic scale and $Z(\tau)$ being given by Eq. (6). (ii) The thrust function $\mathcal{T} = \pm(1 + \epsilon)$ where the sign gives the direction of acceleration and where $\epsilon$ is the reflectivity of the sail. A perfectly absorbing sail, a *black* one, exhibits $\epsilon = 0$, while a perfectly reflecting *white* one has $\epsilon = 1$. For the emission rockets, we can assume i) that the output power is constant [see solutions (15) and (16) with $\mathcal{P} = -1$] or ii) that the output power decays with proper time, $\mathcal{P}(\tau) = -\exp(-s/S)$, with $S$ being the power decay time, as would happen if the emission rocket was powered by the radiative cooling of some radioactive material. In these cases (i) and (ii), $\mathcal{T} = \pm 1$. Figures 1–4 present the kinematics of these four radiation rockets.

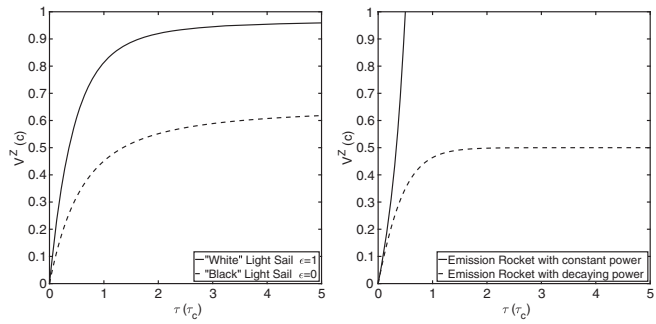

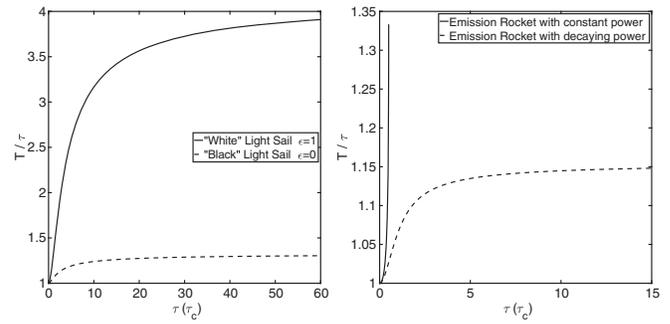

FIG. 1. Evolution of the 3-velocity $V^Z$ with respect to the inertial observer $\mathcal{O}$ with the traveler's proper time $\tau$ for absorption (left) and emission (right) rockets.

FIG. 3. Time dilation effect, shown as the ratio of duration for inertial observer $\mathcal{O}$ to traveler $\mathcal{O}'$'s proper time $T/\tau$, for absorption (left) and emission (right) rockets.

The evolution of a rocket's velocity, starting from rest, and of the rocket's inertial masses as the traveler's proper time evolves are given in Figs. 1 and 2, respectively. The light sails will only reach some fraction of the speed of light $c$ asymptotically, while their inertial masses eventually freeze. The emission rocket with constant driving power reaches $c$ after some finite proper time, at which its inertial mass vanishes. Once $c$ is reached, the proper time of the traveler freezes and ceases to elapse. If the driving power is decaying exponentially with time, the emission rocket asymptotically reaches only some fraction of $c$, while its inertial mass finally freezes. In the figures, we have used the following value of the power decay time $S = 1/3$.

The acceleration of the rockets causes time dilation for the travelers as illustrated in Fig. 3. Light sails reach asymptotically a time dilation factor $T/\tau$ of around 4 for $\epsilon = 1$ (white sail) and around 1.25 for $\epsilon = 0$ (black sail). The emission rocket with constant power formally reach an infinite time dilation factor after some finite proper time since, as soon as it has reached $c$, proper time of the traveler freezes, and the ratio $dT/d\tau \to \infty$. In the case of decaying internal power, the emission rocket finally experiences a frozen time dilation factor, around 1.15 for $S = 1/3$ as can be seen from Fig. 3.

Finally, we give in Fig. 4 a *Tsiolkovsky* diagram showing the change in velocity with the variation of inertial mass of the radiation rockets. The black light sail with $\epsilon = 0$ and the emission rocket both satisfy the relativistic Tsiolkovsky equation (14), while the Tsiolkovsky curve for the white sail with $\epsilon = 1$ moves from close to the purely absorbing case $\epsilon = 0$ to the emission rocket case. It is important to bear in mind that this mass loss is a purely relativistic effect due to the interaction of the rocket with radiation and is therefore quite different than the reaction process due to the ejection of massive propellant.

Actually, there exists a well-known analytical solution that applies to radiation rockets as well. This special case is the so-called hyperbolic motion (cf. Refs. [5,14]), i.e., relativistic motion with a constant norm of the 4-acceleration $\dot{\lambda}^\mu$, with units m$^{-1}$, $|\dot{\lambda}^\mu \dot{\lambda}_\mu| = a^2 \geq 0$. For the characteristic time, we therefore choose $\tau_c = 1/(c|a|)$ so that $\psi = \psi_0 + d.f.(s-s_0)$ and $\mathcal{M} \equiv m/m_0 = \exp(f.(s-s_0))$ with $f = \pm 1$ giving the rocket type ($+1$ absorption; $-1$ emission) and $d = \pm 1$ so that the sign of the thrust is $f.d$. The corresponding 4-force components can now be deduced from Eqs. (12): $f^T = (m_0 c^2/\tau_c).f.\exp(2f(s-s_0))$ and $f^Z = d.f^T$. In this solution, the speed of light $c$ is reached asymptotically (when $s \to \infty$), while the rocket's inertial mass becomes exponentially large (absorption rocket) or decays exponentially (emission rocket). In terms of the above examples, the emission rocket with a decay time $S = 1/2$ precisely corresponds to this solution of the constant norm of 4-acceleration.

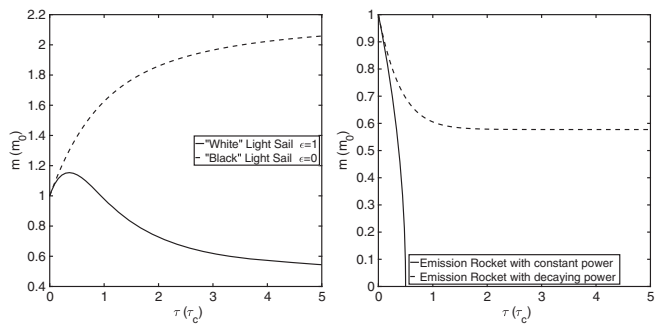

FIG. 2. Evolution of the rocket's inertial mass $m$ with the traveler's proper time $\tau$ for absorption (left) and emission (right) rockets.

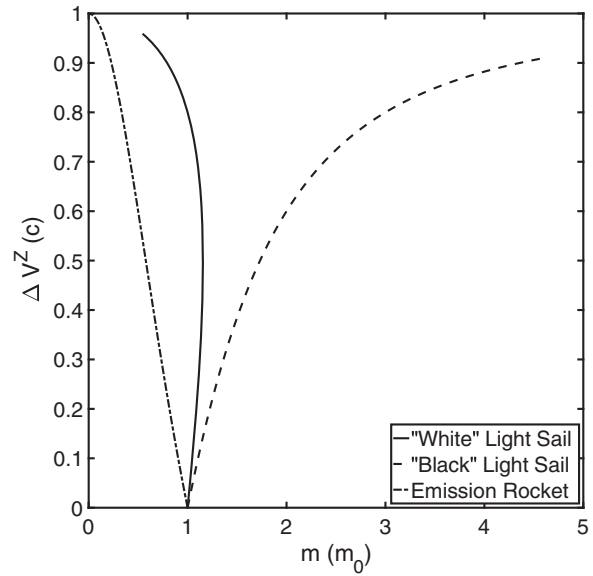

FIG. 4. Tsiolkovsky diagram of the variation of velocity as a function of the rocket mass.

### C. Applications to interstellar travels

#### 1. Acceleration phase of a light sail

We start by applying our modeling to the starshot project [4] in which tiny probes attached to the light sail will be accelerated by high power laser shots from the ground to reach a relativistic velocity after a short acceleration phase before heading to Proxima Centauri for a flyby. We assume here a mass of 10 g for the probe and the sail and the power of the laser beam at the source of $P = 100$ GW, decaying with distance as the inverse of the distance to the source as in the previous section. Figure 5 gives the evolution of the 3-velocity [with Newtonian result $V = P/(cm_0).(1+\epsilon)\tau$ indicated by dashed lines], the 3-acceleration (the last two with respect to inertial observer), and the mass with respect to the proper time $\tau$ during about one hour of continuous push by the lasers. Two different values of the reflexivity $\epsilon = 0$ and $\epsilon = 1$ are indicated, to give an idea of the spread of these kinematical variables with the reflexivity. The total energy cost of the mission roughly corresponds to the amount of energy spent by the power source during the whole acceleration phase, $E = P \times \tau$ (for constant power of the source), which corresponds to $100\,\text{GW} \times 4000\,\text{s} \approx 10^{14}\,\text{J}$.

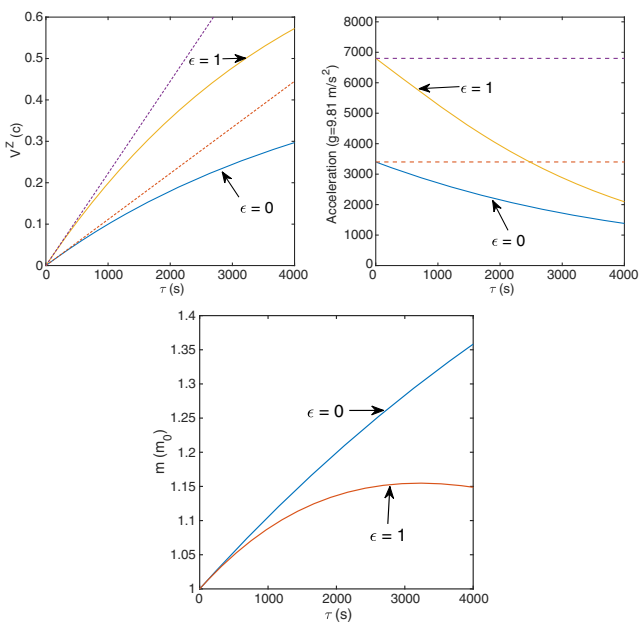

FIG. 5. Acceleration phase of a laser-pushed light sail, with a beam power at a source of 100 GW and a mass of 10 g for reflexivity $\epsilon = 0$ and $\epsilon = 1$. Upper plots: velocity (left) and acceleration (right) with respect to inertial observer; bottom plot: mass variation.

In about an hour of continuous propulsion, the 10 g probe reaches a velocity between 0.3 and approximately $0.6 \times c$ for corresponding acceleration decreasing from the range $[6000; 3000] \times g$ to $[2000; 1000] \times g$. This decrease of the acceleration is a purely relativistic effect. Finally, the mass relative variation of the probe lies in the range 15%–35%, which is non-negligible for performing corrections of trajectory with the embarked photon thrusters.

#### 2. Traveling to Proxima Centauri with an emission radiation rocket

The next application of our former results is a simple modeling of interstellar travels to Proxima Centauri, located about 4 light years away, with large emission radiation rockets. This example is purely illustrative, and we will not list the numerous engineering challenges that must be overcome in order to even start thinking about such a mission, yet it will clearly show the major impediment of interstellar travel: the energy cost.

Let us consider a model of a rocket propelled by the redirection of the blackbody radiation emitted by a large hot surface in radiative cooling. The power driving the rocket is therefore given by $P = \sigma A T^4$ where $\sigma = 5.670373 \times 10^{-8}\,\text{W}/(\text{m}^2 \cdot \text{K}^4)$ is the Stefan-Boltzmann constant and $A$ is the surface of the radiator at temperature $T$. To fix the ideas, we choose the total mass of the rocket to be $m_0 = 100$ tons. We also assume a decay time of $\mathcal{S} = \tau_c$ with $\tau_c = m_0 c^2/P$ being the characteristic timescale.

Figure 6 presents the evolution of several interesting kinematical quantities for both a return and a single trip at a distance of approximately 4 light years. The presented single trip is done with a radiator of $A = 1\,\text{km}^2$ at temperature $T \sim 8.4 \times 10^3$ K (and a total driving power $P \sim 3 \times 10^2$ terawatts) while the presented return trip is done with $A = 100\,\text{km}^2$ at $T = 3000$ K for a power source of $P \sim 4 \times 10^2$ terawatts. Those parameters have been chosen for illustrative reasons and do not pretend to be feasible; simply remember that the characteristic power of a civil nuclear reactor is of order 1 GW. In the top left plot of Fig. 6, one can see the velocity pattern of the trajectories. In the single trip, the rocket first accelerates at a speed of approximately $0.94 \times c$ before it must be turned upside down for deceleration after about 8 months and finally arrives at destination in about 3 years. The acceleration with respect to inertial observers is not shown but is not above $1.25 \times g$. In the case of the return trip, the rocket reaches about $0.97 \times c$ after 5 months of acceleration before flipping for deceleration and reaches the destination after approximately 2.7 years. However, the rocket does not stop and immediately goes back toward its departure location. After having reached a return velocity of about $-0.97 \times c$ at mission time of approximately 3 years, the returning rocket will have to flip for deceleration and finally arrive back at home at rest after approximately 5.4 years. However, the total duration of the return trip from the point of view of the inertial observer who stayed home is approximately 10.6 years, as a consequence of time dilation (see Fig. 6 lower right panel). Similarly, the single trip has been performed in about 3.5 years from the point of view of the traveler but about 6 years from the point of view of his home. The accelerations underwent by the return rocket are less than 2g's. In the upper right panel of Fig. 6, one can see the mass decrease of the rockets as a function of mission time. The flips of the rockets have been assumed instantaneous, which explains the shape of the curves around the flips. Worldlines of the interstellar travels in inertial coordinates are shown in the lower left panel of Fig. 6. The worldline of the charterer, i.e., the observer who stayed home corresponds to the $(Z = 0, T)$ vertical line. The events of departure and arrival of the return rocket are given by the intersection of the traveler's worldline and the vertical axis.

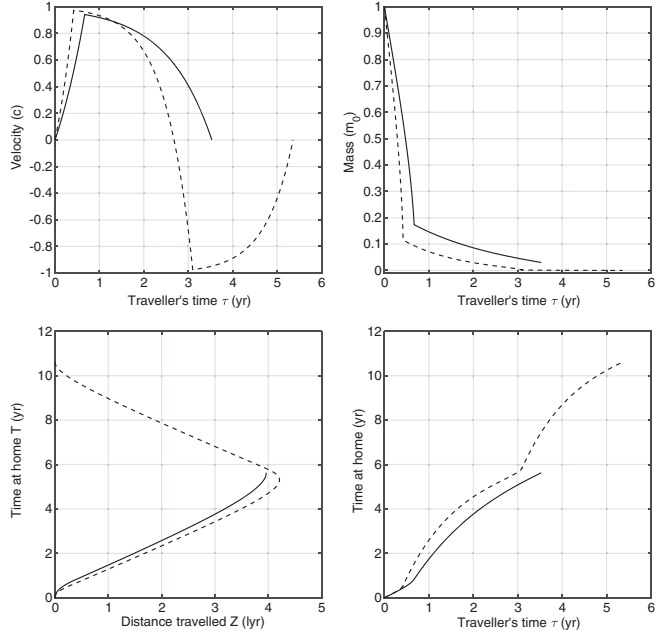

FIG. 6. Single and return trips (respectively, straight and dashed lines) toward any destination located about 4 light years away, like the Alpha Centauri star system, with an emission radiation rocket. Upper left: velocity as a function of traveler's proper time $\tau$; upper right: mass variation of the photon rocket during the trip; lower left: worldline $\mathcal{L}$ of the trips in inertial coordinates; lower right: time dilation for the traveler.

We finish this section by echoing our Introduction. Emission radiation rockets do not involve new physics and actually are propelled by the least noble sort of energy, which is heat. By maintaining acceleration throughout the trip, one can significantly reduces its duration by comparison of the one measured by an observer who stayed at the departure location. However, the major impediment is that such rockets need to embark an extreme power source, 100 terawatts, within the lowest mass possible, and make it work all along the several years of travel duration. By integrating $f^T(\tau)$ all along the trips, one finds an estimation of the total energy cost $E$ of an interstellar mission to Proxima Centauri with a 100 tons scale spaceship is the unbelievable figure of $E \sim 9 \times 10^{21}$ J. To fix the ideas, this amount of energy corresponds to about 15 times the world energy production in 2017. Needless to say, no one can (presently?) afford such interstellar travel, and one should instead rely on another scheme, like the starshot concept [4], to physically investigate nearby star systems.

## III. PROPAGATION OF LIGHT TOWARD OBSERVERS ABOARD RADIATION ROCKETS

### A. Geodesics in the Hamiltonian formalism

We now investigate spacetime geometry around the radiation rockets through characterizing null geodesics, which are nothing but the trajectories of light. As explained in Sec. II, Kinnersley's spacetime geometry (3) is specified through the functions $M(\tau) = 2Gm(\tau)/c^2$ (with dimension length) and $\alpha = -\dot{\psi}/c$ (with dimension inverse of length) of proper time $\tau$ [which is also the value of the null coordinate $u$ in Eq. (3) in the limits $M \to 0$ or $r \to \infty$]. These functions $M$ and $\alpha$ are associated, respectively, to the mass and to the 4-acceleration (i.e., $|g_{\mu\nu}\ddot{\lambda}'^{\mu}\ddot{\lambda}'^{\nu}| = \alpha^2$) of the rocket. Light rays incoming toward (or outgoing from) the traveler must be computed from the retarded (advanced) metric. Any geodesic curve $\mathcal{G}$ is specified in these coordinates by the following set of functions $\mathcal{G} = (c.\tau(\sigma), r(\sigma), \theta(\sigma), \varphi(\sigma))$ with $\sigma$ some affine parameter on the geodesic. These functions are solutions of the geodesic equation,

$$\frac{d^2 x^\gamma}{d\sigma^2} + \Gamma^\gamma_{\alpha\beta} \frac{dx^\alpha}{d\sigma} \frac{dx^\beta}{d\sigma} = 0, \quad (17)$$

but those equations are difficult to handle in the metric (3), as can be seen in Refs. [14,16]. This is why we prefer here to proceed with the so-called Hamiltonian formulation of geodesics [21].

Geodesic equations (17) are Euler-Lagrange equations for the action of a pointlike particle $S = \int m.ds$ (with $ds$ being the line element in spacetime) but also of the following Lagrangian: $L = 1/2 g_{\mu\nu} \dot{x}^\mu \dot{x}^\nu$ with $\dot{x}^\mu = \frac{dx^\mu}{d\sigma}$. Introducing canonical momenta as usual by $p_\alpha = \frac{\partial L}{\partial \dot{x}^\alpha}$, one can introduce an associated Hamiltonian,

$$H = \frac{1}{2} g^{\alpha\beta} p_\alpha p_\beta. \tag{18}$$

Then, instead of solving Euler-Lagrange equations (17), one can advantageously solve rather their Hamiltonian counterparts:

$$\begin{cases} \frac{dx^\mu}{d\sigma} = g^{\mu\nu} p_\nu \\ \frac{dp_\mu}{d\sigma} = -\frac{1}{2} \frac{\partial g^{\alpha\beta}}{\partial x^\mu} p_\alpha p_\beta. \end{cases} \tag{19}$$

The contravariant metric components $[g^{\alpha\beta}]$ read

$$\begin{pmatrix} 0 & \pm 1 & 0 & 0 \\ \pm 1 & 2r\alpha\cos(\theta) - 1 + 2M/r & -\alpha\sin(\theta) & 0 \\ 0 & -\alpha\sin(\theta) & -1/r^2 & 0 \\ 0 & 0 & 0 & -1/(r^2.\sin^2(\theta)) \end{pmatrix},$$

where the $+$ ($-$) sign is for incoming (outgoing) geodesics. According to this, the (co)geodesic equations (19) can now be written down ($c = 1$, a dot indicating a derivative with respect to $\tau$),

$$\frac{d\tau}{d\sigma} = \pm p_r \tag{20}$$

$$\frac{dr}{d\sigma} = \pm p_\tau - p_r \left(1 - 2\frac{M}{r} - 2\alpha r \cos(\theta)\right) - p_\theta \alpha \sin(\theta) \tag{21}$$

$$\frac{d\theta}{d\sigma} = -\frac{p_\theta}{r^2} - p_r \alpha \sin(\theta) \tag{22}$$

$$\frac{d\varphi}{d\sigma} = -\frac{p_\varphi}{r^2 \sin^2(\theta)} \tag{23}$$

$$\frac{dp_\tau}{d\sigma} = p_r p_\theta \sin(\theta) \dot{\alpha} - \frac{p_r^2}{r}(\dot{\alpha}\cos(\theta) r^2 + \dot{M}) \tag{24}$$

$$\frac{dp_r}{d\sigma} = \frac{p_r^2}{r^2}(-\alpha\cos(\theta) r^2 + M) - \frac{p_\varphi^2}{r^3 \sin^2(\theta)} - \frac{p_\theta^2}{r^3} \tag{25}$$

$$\frac{dp_\theta}{d\sigma} = p_r \alpha (p_\theta \cos(\theta) + p_r r \sin(\theta)) - \frac{p_\varphi^2 \cos(\theta)}{r^2 \sin^3(\theta)}, \tag{26}$$

with $p_\varphi$ a constant of motion, since the metric does not explicitly depend on $\varphi$ (axial symmetry). The corresponding Hamiltonian, which is also a constant of motion $dH/d\sigma = 0$, is given by

$$H = -\frac{p_\varphi^2}{2r^2 \sin^2(\theta)} - \frac{p_r^2}{2}\left(1 - \frac{2M}{r} - 2r\alpha\cos(\theta)\right)$$
$$- \alpha\sin(\theta) p_r p_\theta \pm p_\tau p_r - \frac{p_\theta^2}{2r^2}. \tag{27}$$

For light rays, or null geodesics, $H$ identically vanishes, while for matter geodesics, $H < 0$, and both types of geodesics obey the same set of ordinary differential equations (19).

A first trivial particular solution is given by constant $\tau$, $\theta$, and $\varphi$, while $r \sim \sigma$ ($p_r = p_\theta = p_\varphi = 0$, $p_\tau = $ cst), which shows that $\tau$ is indeed a null coordinate. Special relativity describes motion at constant velocity and vanishing mass corresponding to the special case $M = \alpha = 0$, which yields a second class of particular solutions: these are null geodesics, $p_\varphi = p_\theta = 0$, $p_r = 2p_\tau \neq 0$, and therefore $\tau = \tau_0 \mp 2r$ (the case of vanishing $p_r$ is the previous trivial solution).

In the general case, spacetime geometry around the photon rocket is ruled by the two functions $M$ and $\alpha$, which can be obtained by solving the relativistic rocket equations given the radiation reaction 4-force (see Sec. II). However, the mass function can be safely neglected in any physical situation except those of huge *luminosity* of the rocket. To see this, one can simply rewrite the metric (3) with the characteristic units introduced before, by setting $s = \tau/\tau_c$ and $R = r/(c\tau_c)$ with $\tau_c = m_0 c^2/|f^T|$ being the characteristic timescale of the photon rocket physical system. Doing so, the mass term $M/r$ in Eq. (3) reduces to $\mathcal{M}/R \times (G|f^T|/c^5)$ ($m = \mathcal{M}.m_0$), and the product $\alpha \cdot M = -\frac{d\psi}{ds}.\mathcal{M} \times (G|f^T|/c^5)$. This means that one can safely neglect the mass effect carried by $M$ in front of the acceleration effect due to $\alpha$ as long as

$$\frac{GL}{c^5} \ll 1, \tag{28}$$

where we have replaced $|f^T|$ by $L$, the luminosity driving the photon rocket. It is surprising to notice that it is not the rest mass $m_0$ but the luminosity $L$ that matters for the photon rocket spacetime geometry [22]. One can interestingly ask for which kind of physical phenomena the mass function $M$ should not be neglected anymore. Well, a remarkable example is binary black hole mergers and their recoil (also dubbed a black hole kick) through the anisotropic emission of gravitational waves during merging. For instance, in the very final moments of the GW150914 binary black hole merger event, the emitted power of gravitational radiation reached about $10^{49}$ W [23], so that $GL/c^5 \approx 10^{-3}$. With such luminosity, the complete photon rocket metric, including $M$ and $\alpha$, should be considered in characterizing the spacetime geometry around a black hole merger self-accelerated by its anisotropic emission of gravitational waves. As a matter of comparison, electromagnetic record luminosities are far lower: the most brilliant supernova reached a luminosity of *only* $10^{38}$ W [24] (with $GL/c^5 \approx 10^{-15}$), while the brightest quasar, 3C273, has luminosity of order $10^{39}$ W [25] (yielding $GL/c^5 \approx 10^{-14}$). It is also worth noticing that the luminosity $L = c^5/(2G)$ appears as an absolute upper bound build from dimensional considerations in general relativity by Hogan [26] and is dubbed Planck luminosity.

In what follows, we will apply photon rockets to models of interstellar travel and will assume *weak* luminosities in the sense of Eq. (28) such that their mass function is negligible $M \ll 1$. Future works should investigate further the applications of photon rocket spacetimes to the modeling of astrophysical events such as black hole merger recoil.

### B. Relativistic aberration and Doppler effect for accelerated relativistic travelers

We now derive from null geodesics of the Kinnersley metric two important effects on the light signals received by an accelerated observer moving at relativistic velocities: the deviation of the incidence angle, also called as relativistic aberration, and the frequency shift. We assume the traveler undergoes an accelerated trajectory, starting from rest at $\tau = 0$ and $Z(0) = T(0) = \psi(0) = 0$ such that the coordinates $(\theta, \varphi)$ at start $\tau = 0$ are usual spherical coordinates (see Eqs. (4) and Ref. [9]) that can be used to map the reference celestial sphere. This reference celestial sphere also corresponds to the one of the inertial observer of which the worldline is tangent to the traveler's worldline at departure $\tau = 0$. We are interested in the trajectories of light rays between departure $\tau = 0$ up to their reception by the interstellar traveler at some proper time $\tau = \tau_R$, since the paths of light rays before traveler's departure $\tau < 0$ are not affected by its motion (the traveler stayed at rest at home at $\tau < 0$). We also assume here that $M = 0$, since we are not considering extreme luminosities as mentioned above, and therefore Eq. (3) will describe Minkowski flat spacetime (see also Ref. [9]) but from the point of view of the accelerated traveler. In this accelerated frame, light rays will undergo angular deviation, leading to relativistic aberration and frequency shifts (Doppler effect) which are different from those described by special relativity with motion at constant velocity. Both effects are of crucial importance for the interstellar traveler since this affects not only its telecommunications but also its navigation by modifying positions and color of the guiding stars. As we shall see below, these effects both depend on the trajectory followed by the traveler.

Indeed, light ray trajectories are solutions of the geodesic equations Eqs. (20)–(26) with a null value of the Hamiltonian (27), and these solutions depend on the time variation of the acceleration function $\alpha(\tau)$ ($M$ can be safely neglected unless one faces extreme luminosities).

Here, we will solve the geodesic equations (20)–(26) by integrating numerically backward in time, from the reception of the light ray by the traveler at ($\tau = \tau_R$, $r = 0$, $\theta = \theta_R$, $\varphi = \varphi_R$) back to the time of the traveler's departure $\tau = 0$ at which the light ray was emitted by the reference celestial sphere at $\tau = 0$, $r = r_E$, $\theta_E$, $\varphi_E$. In order to compute the local celestial sphere of the traveler (who is located at $r = 0$), we are interested in incoming light rays with the two following features. First, they have a null impact parameter (i.e., $p_\varphi = 0$ yielding $\varphi_R = \varphi_E$ by Eq. (23). Second, smoothness of the null geodesics at reception $r(\tau_R) = 0$ requires that $p_\theta(\tau_R) = 0$. Since $\tau_R$ and $\theta_R$ are considered as free parameters, this leaves only two initial conditions, $p_r(\tau_R)$, $p_\tau(\tau_R)$, to be determined. From Eqs. (20)–(26), we can set, without loss of generality, $p_r(\tau_R) = 1$ so that $\tau \approx \sigma$ at reception (the affine parameter is then simply scaled by choice to proper time at reception). $p_\tau(\tau_R)$ must then be obtained by solving the Hamiltonian constraint $H = 0$ (27) with respect to $p_\tau$, given all the other initial conditions at $\tau = \tau_R$. This achieves fixing our set of initial conditions at given $\tau_R$. Then, integrating backward the geodesics equations (20)–(26) until the rays were emitted from the reference celestial sphere at $\tau = 0$, one can compute the angular deviation and the frequency shifts of light between the reference celestial sphere at $\tau = 0$ and the local celestial sphere of the traveler at $\tau = \tau_R$.

To compute relativistic aberration for the accelerated traveler, one needs to account for two contributions. The first is the angular coordinate change $\theta_R \neq \theta_E$ at both ends of the null geodesic (remember that $p_\varphi = 0$ so that $\varphi_R = \varphi_E$). The second input comes from the fact that the angular coordinate $\theta_R$ does not correspond anymore to the usual spherical coordinate at $\tau = \tau_R$ and $\psi_R \neq 0$. To find the corresponding angle $\Theta_R$ on the local celestial sphere of the traveler, one has to move back to the instantaneous rest frame of the traveler. This is done by imposing

$$Z_R - Z_{\mathcal{L}}(\tau_R) = \rho \cos \Theta_R$$

and

$$X^2 + Y^2 = r_R^2 \sin^2(\theta_R) = \rho^2 \sin^2(\Theta_R)$$

in the coordinate transformation (4), where $\rho$ and $\Theta_R$ are local spherical coordinates [27]. Doing so, we can write down the correspondence relation between both angles of incidence $\Theta_R$ in the local traveler's frame and the angular coordinate $\theta_R$ at reception $\tau = \tau_R$ as

$$\tan \Theta_R \cdot (\beta + \cos \theta_R) = \sin \theta_R \cdot (1 - \beta^2)^{1/2} \quad (29)$$

with $\beta = \tanh \psi_R$. One can check that Eq. (29) is identical to the formula of relativistic aberration in special relativity as obtained by Einstein in Refs. [5,28] for motion at constant velocity for which $\theta_R$ is then the angle of incidence as measured by the observer at rest (and $\theta_E = \theta_R$ in special relativity). The aberration angle for the accelerated observer is therefore given by $\Theta_R = \Theta_R(\theta_E)$ with $\Theta_R$ given by Eq. (29) in which $\theta_R$ is a function of $\theta_E$ as obtained by the backward integration of null geodesic equations.

Of course, in the case of motion at constant velocity $\psi = $ cst, $\alpha = 0$, and null geodesics are given by the trivial solution $p_\varphi = p_\theta = 0$, $p_r = 2p_\tau = 1$, $\tau = \tau_0 \mp 2r$, such that one has $\theta_R = \theta_E$ leading to the special relativistic aberration described by Eq. (29). Quite interestingly, among all possible traveler's worldlines, there is a non-trivial one for which there is also no angular deviation $\theta(\sigma) = $ cst, or in other words $\theta_R = \theta_E$, and the relativistic aberration of the accelerated traveler reduces to the one of special relativity and motion at a constant velocity. This is the case when $\alpha(\tau) = $ cst, which corresponds to the hyperbolic motion of Sec. II. Indeed, setting $d\theta/d\sigma = 0$ in Eq. (22), one obtains that

$$p_\theta = -p_r r^2 \alpha \sin\theta. \tag{30}$$

Then, since $\dot\alpha = \dot M = 0$ ($\alpha = $ cst, $M = 0$), Eq. (24) yields that $p_\tau = $ cst, of which the value can be obtained from the Hamiltonian constraint $H = 0$. Solving Eq. (27) with respect to $p_\tau$ and assuming $p_r \neq 0$, one finds that

$$p_\tau = \frac{p_r}{2}(1 - 2r\alpha\cos\theta - \alpha^2 r^2 \sin^2\theta). \tag{31}$$

Putting Eqs. (30) and (31) and $p_\varphi = 0$ into Eqs. (21) and (25), we obtain

$$\frac{dr}{d\sigma} = -p_\tau \tag{32}$$

$$\frac{dp_r}{d\sigma} = -p_r^2(\alpha\cos\theta + r\alpha^2 \sin^2\theta). \tag{33}$$

Finally, one can use Eqs. (30), (32), and (33) and $p_\varphi = 0$ and $d\theta/d\sigma = 0$ to retrieve Eq. (26), showing that $\alpha = $ cst implies $d\theta/d\sigma = 0$.

Let us now focus on the Doppler effect, i.e., the frequency shift of light signals that are measured by the accelerated traveler in relativistic flight. The energy of the photon measured at spacetime event $e$ by some observer is given by $E_e = h \cdot \nu_e = (p_\mu \lambda'^\mu)_e$ with $\lambda'^\mu$ being the unit tangent vector to the observer $\mathcal{O}'$ s worldline, $h$ being Planck's constant, and $\nu_e$ being the measured frequency of the photon. In this application, the receiver is the traveler, with worldline $(r = 0, \tau)$ in his local coordinates, so that the received frequency of the photon is $\nu_R \propto p_\tau(\tau_R)$. At start $\tau = 0$, we consider an emitter on the reference celestial sphere that has no proper motion with respect to the home position, corresponding to a worldline given by fixed inertial coordinates $(X, Y, Z)_{\tau=0} = (X_E, Y_E, Z_E)$ and proper time $T$. This models a fictitious star located at $(X_E, Y_E, Z_E)$ with assumed no proper motion at the position $(\theta_E, \varphi_E)$ on the reference celestial sphere and of which the light frequencies of the emitted light rays are those observed by the inertial observer stayed at home. The components of the tangent vector to the emitter's worldline are given by $dx^\mu/dT = d(\tau_E, r_E, \theta_E, \varphi_E)/dT$ (and $d\varphi_E/dT = 0$) and must be computed from Eqs. (4):

$$T_E = T_\mathcal{L} + r_E \cdot [\cosh(\psi_E) + \cos(\theta_E) \cdot \sinh(\psi_E)] \tag{34}$$

$$Z_E = Z_\mathcal{L} + r_E \cdot [\sinh(\psi_E) + \cos(\theta_E) \cdot \cosh(\psi_E)] \tag{35}$$

$$(X_E + Y_E)^{1/2} = r_E \sin\theta_E. \tag{36}$$

By differentiating each side of Eq. (36), one finds

$$\frac{d\theta_E}{dT} = -\frac{\tan\theta_E}{r_E}\frac{dr_E}{dT}, \tag{37}$$

which we can substitute into the differentiations of Eqs. (34) and (35) with respect to $T$ together with Eqs. (5) and (6) to find a system of linear equations for the unknowns $(\frac{d\tau_E}{dT}, \frac{dr_E}{dT})$. Solving this system yields

$$\frac{d\tau_E}{dT} = \cosh\psi_E + \cos\theta_E \sinh\psi_E \tag{38}$$

$$\frac{dr_E}{dT} = -\cos\theta_E \sinh\psi_E$$
$$+ \alpha_E r_E \cos\theta_E(\cosh\psi_E + \cos\theta_E \sinh\psi_E). \tag{39}$$

From this result, it is possible to retrieve the Doppler shift formula of special relativity since

$$E_E = p_\tau(0)\frac{d\tau_E}{dT} + p_r(0)\frac{dr_E}{dT}$$
$$= \frac{1}{2}(-\cos\theta_E \sinh\psi_E + \cosh\psi_E)$$

(remind that $p_\tau = 1/2 = p_r/2$, $p_\theta = p_\varphi = 0$, $\alpha = 0$ in this case) and

$$E_R = p_\tau(\tau_R)\frac{d\tau_R}{dT} + p_r(r_R)\frac{d\tau_R}{dT} = 1/2$$

(since $\frac{d\tau_R}{dT} = 1$ and $\frac{dr_R}{dT} = 0$). The ratio $E_E/E_R$ then identically matches the relativistic Doppler effect formula (cf. Ref. [4]),

$$\left(\frac{\nu_R}{\nu_E}\right)_{\text{SR}} = (1 - \beta_E \cos\theta_E)^{-1} \cdot \sqrt{1 - \beta_E^2}, \tag{40}$$

with $\beta_E = \tanh\psi_E$. For accelerated motions, $\alpha_E \neq 0$, and considering that we started at rest, $\psi_E = 0$, Eqs. (37)–(39) yield simply

$$\frac{d\tau_E}{dT} = 1$$
$$\frac{dr_E}{d\tau} = r_E \cos(\theta_E)\alpha_E$$
$$\frac{d\theta_E}{d\tau} = -\sin(\theta_E)\alpha_E.$$

From these relations, one can compute the unit tangent vector of the emitter at time of emission,

$$\lambda'^{\tau,r,\theta} = \frac{1}{N}\frac{d(\tau_E, r_E, \theta_E)}{dT},$$

where

$$N = \sqrt{g_{\alpha\beta}\frac{dx^\alpha}{dT}\frac{dx^\beta}{dT}}$$

is the norm of the tangent vector $\lambda'^\mu = dx^\mu/dT$.

The photon energy at emission $e = E$ is therefore $E_E = (p_\mu \cdot \lambda'^\mu)|_{(\tau_E=0)}$. Finally, the frequency shift is simply given by $E_R/E_E$, the ratio of the received frequency $\nu_R$ over the emitted one, $\nu_E$.

The first type is (a) the hyperbolic motion, given by $\psi = s$ ($\alpha' = -d\psi/ds = -1$ in characteristic units) and $\mathcal{M} = e^{\pm s}$ with $s = \tau/\tau_c$, $\tau_c$ being the characteristic time defined in Sec. II. This case of hyperbolic motion is a critical point for which aberration of the accelerated traveller is the one described by special relativity. For reasons that will appear clearly below, we choose to consider also two other types: (b) a perfect light sail (see Sec. II) for which $\alpha'$ monotonically increases from $\alpha'(0) = -2$ to $\alpha' \to 0$ when $s \to \infty$ and (c) an emission rocket with constant output power Eqs. (15) and (16) in which $\alpha' \le -1$.

Figure 7 presents the convergence of null geodesics toward the observer at [$\tau = \tau_R$, $r = 0$ and $V^Z(\tau_R) = 0.9 \cdot c$] as plots of the proper time $\tau(\sigma)$ along the null geodesics as functions of $r \cdot \cos\theta$ for the three different photon rockets mentioned above hyperbolic motion, light sail, and emission rocket. The different curves corresponds to different initial conditions $\theta_R$. One can clearly see the Minkowskian regime $\tau \sim 2r$ of the null geodesics as $r(\sigma) \to 0$, and the metric (3) becomes close to the case of motion at constant velocity $\alpha = 0$ (remember that $M = 0$ here). Also shown in Fig. 7 is a consistency check through the violation of the Hamiltonian constraint $H = 0$ along the null geodesics; this constraint is found to be pretty stable, and the value of $H$ is kept around the order of magnitude of the numerical integrator tolerance. In establishing the following results, we have always monitored this Hamiltonian constraint, which has been found to be quite robust and always controlled by the numerical integration tolerance.

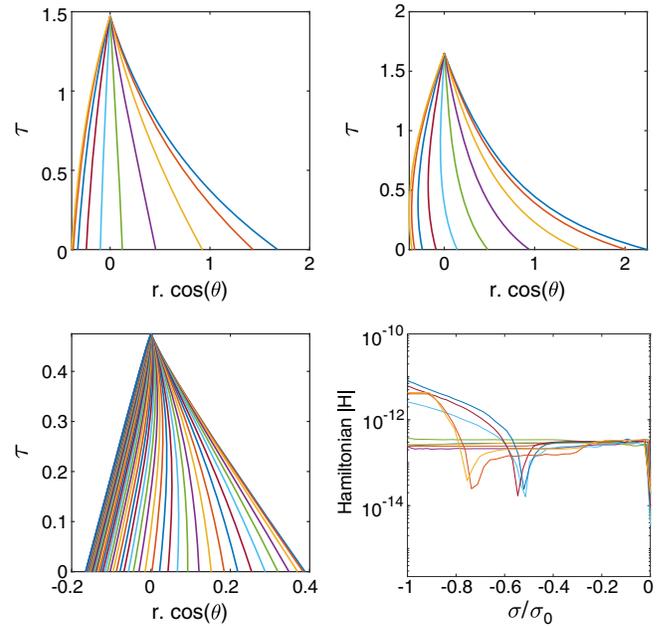

FIG. 7. Convergence of null geodesics toward the traveler at ($r = 0$, $\tau = \tau_R$) and $\tanh(\psi_R) = 0.9$ for different values of the initial emission angle $\theta_E$ (upper left panel: hyperbolic motion with $\alpha = 1$; upper right panel: light sail; lower left panel: emission rocket with constant driving power; lower right panel: Hamiltonian for the case of hyperbolic motion and a tolerance of the integrator of $10^{-12}$, the exact value for H is zero for null geodesics).

Figure 8 presents the relativistic aberration for the accelerated traveler as proper time evolves aboard a rocket with $\alpha = 1$ (top panel), a light sail (central panel), and an emission photon rocket (bottom panel). The case of hyperbolic motion gives rise to the same relativistic aberration as if the traveler were in motion at constant velocity, $\Theta_R = \Theta_{SR}$, as explained above (see Fig. 8, top panel) and is shown for reference of the two other cases. In the case of a light sail (central panel in Fig. 8), one has that the received angle of incidence $\Theta_R$ gets greater and greater than the angle of incidence at start $\theta_E$ as the traveler accelerates and increases its velocity. However, one can see from Fig. 8 (central panel) that this effect is quantitatively smaller than the aberration in special relativity ($\theta_E(\Theta_{SR})$ is shown as a dashed line) for the light sail: $\Theta_R < \Theta_{SR}$. In the case of an emission rocket in Fig. 8 (bottom panel), the angle of incidence as measured by the accelerated traveler $\Theta_R$ is greater than the case of special relativity: $\Theta_R > \Theta_{SR}$. Hence, we have shown that relativistic aberration of an accelerated observer depends on the type of photon rocket.

Figure 9 presents the Doppler effect for the accelerated traveler aboard the three different photon rockets discussed here, through the ratio $\nu_R/\nu_E$ as a function of the direction of reception $\Theta_R$. The unit circle marks the transition from redshifts $\nu_R/\nu_E < 1$ (for directions of reception $\Theta_R$ far from zero) to blueshifts $\nu_R/\nu_E > 1$ (for directions of reception $\Theta_R \approx 0$ close to that of motion). The emission rocket presents Doppler shifts that are found close to those described by special relativity [Eq. (40) for $\beta = 0.95$ is shown as dashed lines in Fig. 9]. For hyperbolic motion and light sails, the Doppler effect for accelerated travelers departs farther and farther from the special relativistic value (40) as the velocity increases. The departures from the special relativistic case are, however, stronger with higher velocities and can be understood since in all the cases considered here we have $\alpha_E \neq 0$. However, it must be noticed that on-axis Doppler shifts (for $\Theta_R = 0; \pi$) are given by the formula from special relativity (40) even when $\alpha_E \neq 0$. Indeed, for $\theta = 0; \pi$, we have that $p_\theta$ is conserved [since $\sin \theta = 0$ in Eq. (22)] and therefore $\theta_E = \theta_R = 0; \pi$ and, from Eq. (29), $\Theta_R = \theta_R = 0; \pi$.

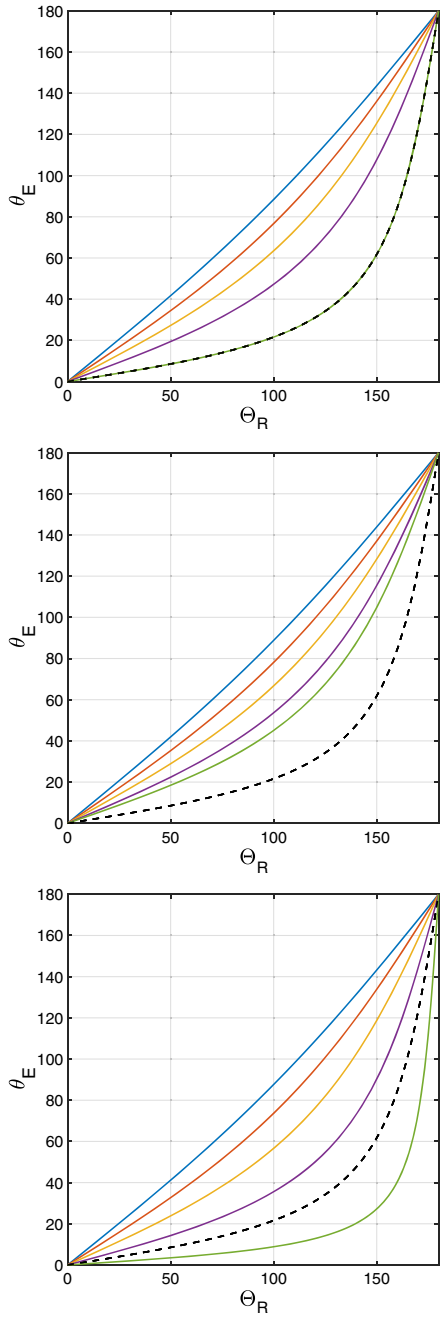

FIG. 8. Relativistic aberration for the accelerated traveler through the angle of incidence at emission $\theta_E$ as a function of the received angle of incidence $\Theta_R$ for the hyperbolic motion (top), light sail (center), and emission rocket (bottom) at various velocities ($V^Z/c = 0.2, 0.39, 0.57, 0.76, 0.95$). Relativistic aberration for a motion at constant velocity of $V^Z/c = 0.95$ in special relativity is given as a black dashed line.

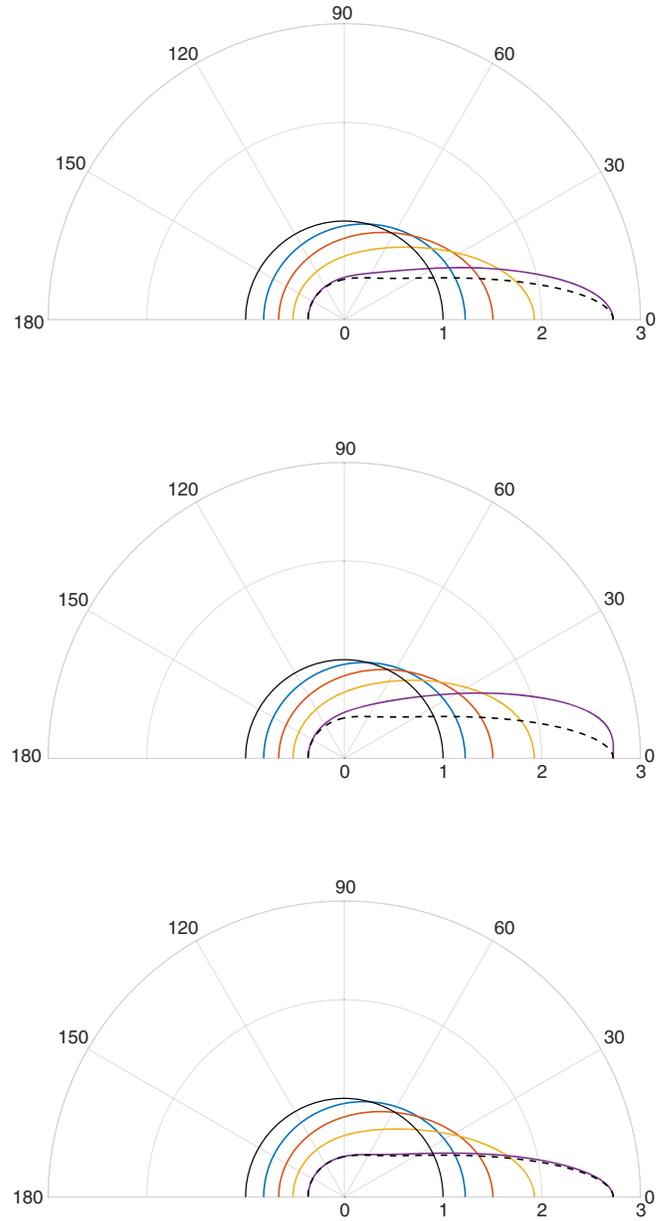

FIG. 9. Directional plots of the frequency shifts $(\nu_R/\nu_E)(\Theta)$ for the hyperbolic motion (top), light sail (center), and emission rocket (bottom) at various velocities ($V^Z/c = 0.2, 0.39, 0.57, 0.76, 0.95$). The Doppler effect for a motion at constant velocity of $V^Z/c = 0.95$ in special relativity is given as a black dashed line, while the unit circle separates redshifts $\nu_R/\nu_E < 1$ from blueshifts $\nu_R/\nu_E > 1$

### C. Application: Deformation of the interstellar traveler's local celestial sphere

In this section, we build a model of the deformation of the local celestial sphere of the accelerated traveler during his trip toward a distant star under the combined effects of relativistic aberration, Doppler frequency shifts, and focusing of light under time dilation.

Our reference celestial sphere will be given by data from the fifth edition of the Yale Bright Star Catalogue [29] in which we choose some star for the traveler's destination and map through appropriate axis rotations the right ascension and declination coordinates onto spherical coordinates $(\theta_E, \varphi)$ at $\tau = 0$ with the axis $Z$ pointing toward the destination star. For each star in the catalog, we can also obtain the temperature from its B-V magnitude[2] from the results in Ref. [30]. With this temperature in hand, we have a blackbody spectrum for each star in the catalog, and from this spectrum, we can associate a specific color from colorimetric considerations [31]. For aesthetic reasons, we choose the destination star as Alnilam, at the center of the Orion belt, and will show only some field of view centered around the front and rear directions of the interstellar rockets. To reconstruct a local view of the accelerated traveler at proper time $\tau_R$, we will loop on each star in the catalog and compute both its local position and color, taking into account relativistic aberration and the Doppler effect as follows. From the angle of incidence $\theta_E$ of a given star in the catalog, we can obtain the observed angle of incidence $\Theta_R$ from our previous results in Fig. 8 and hence the associated position on the traveler's local celestial sphere. The observed angle of incidence $\Theta_R$ of the star will also determine its frequency shift from relations shown in Fig. 9 and a rendering of some star's observed color by applying the associated Doppler shift to the star's blackbody spectrum to obtain a RGB triplet from our colorimetric functions. Finally, each star lying in the field of view is plotted as a sphere located at the found position, with color associated to the shifted blackbody spectrum. We also have to take into account the focusing effect of special relativity: the luminosity of the stars measured by the traveler is increased by the Lorentz factor $\Gamma = (1 - \beta^2)^{-1/2}$ due to time dilation. Therefore, the apparent magnitude for the traveler $m_T$ is related to the magnitude on the reference celestial sphere $m_0$ by the following relation:

$$m_T = m_0 - 2.5 \log_{10} \Gamma.$$

Finally, each star lying in the field of view is plotted as a sphere located at the found position, with color associated to the shifted blackbody spectrum and a size inversely proportional to its visual magnitude $m_T$, which is restricted to $m_T \leq 6$. The accelerated observer is located at the center of the local celestial sphere looking either in the front (Fig. 10) or the rear (Fig. 11) directions of motion.

The evolution of the traveler's celestial sphere heading toward Alnilam as its velocity increases is given in Figs. 10 and 11 for an emission rocket and a light sail. Remember that the acceleration modifies both relativistic aberration and the Doppler effect compared to their descriptions in special relativity. Doing so, the relativistic beaming is observed in the front view (Fig. 10), but it is stronger for the emission rocket and weaker for the light sail than what is predicted by special relativity. One can see this by looking at how the asterism of the Winter Hexagon shrinks more rapidly aboard an emission rocket than aboard a light sail as velocity increases. One can also see how the Big Dipper in the upper left appears earlier in the field of view for the emission rocket than for the light sail (Fig. 10, central panel). The Doppler effect is responsible for the nice reddening of stars outside of some cone centered on the destination, while stars that are observed close to the front direction appear bluer than they are in their rest frame. The solid angle in which stars are bluer is slightly smaller for the light sail than for the emission rocket, but the blueshift in the light sail case is stronger (see Fig. 10, bottom panel, and Fig. 9, top and bottom panels). Figure 11 presents the evolution of the rear view from the photon rockets as time evolves. At $\beta = 0.2$, the rear views are pretty similar aboard both photon rockets, where one can easily recognize the majestic constellations of Scorpius and Sagittarius (bottom of the field of view), Lyra (upper left), Aquila and Delphinus (on the left). Then, as the velocity increases, this rear view is emptied of stars more rapidly aboard the emission rocket than aboard a light sail. This can be seen while looking at how Scorpius and Aquila are leaving the field of rear view. This difference is due to the stronger relativistic aberration for the emission rocket. The Doppler shift on the rear is in both cases very close to the one predicted by special relativity for angles of incidence above 120 deg, as can also be seen from Fig. 9.

From the results presented here, it is possible to model the local celestial sphere of any accelerated relativistic observer, in any journey toward any star neither on a single nor on a return trip. As shown above, the mass function $M$ can safely be neglected for sub-Planckian luminosities. Therefore, one simply needs to provide a smooth acceleration profile $\alpha(\tau)$ for the journey of interest, so that the spacetime metric is well defined without discontinuities, and use our procedure to compute the relativistic effects on incoming light signals.

---

[2]In this context, B and V refers to two different spectral bands in the visible spectrum, which are standard in astrophysics. The B (for Blue) band is centered around a wavelength of 442 nm while the V (for Visible) is centered around 540 nm.

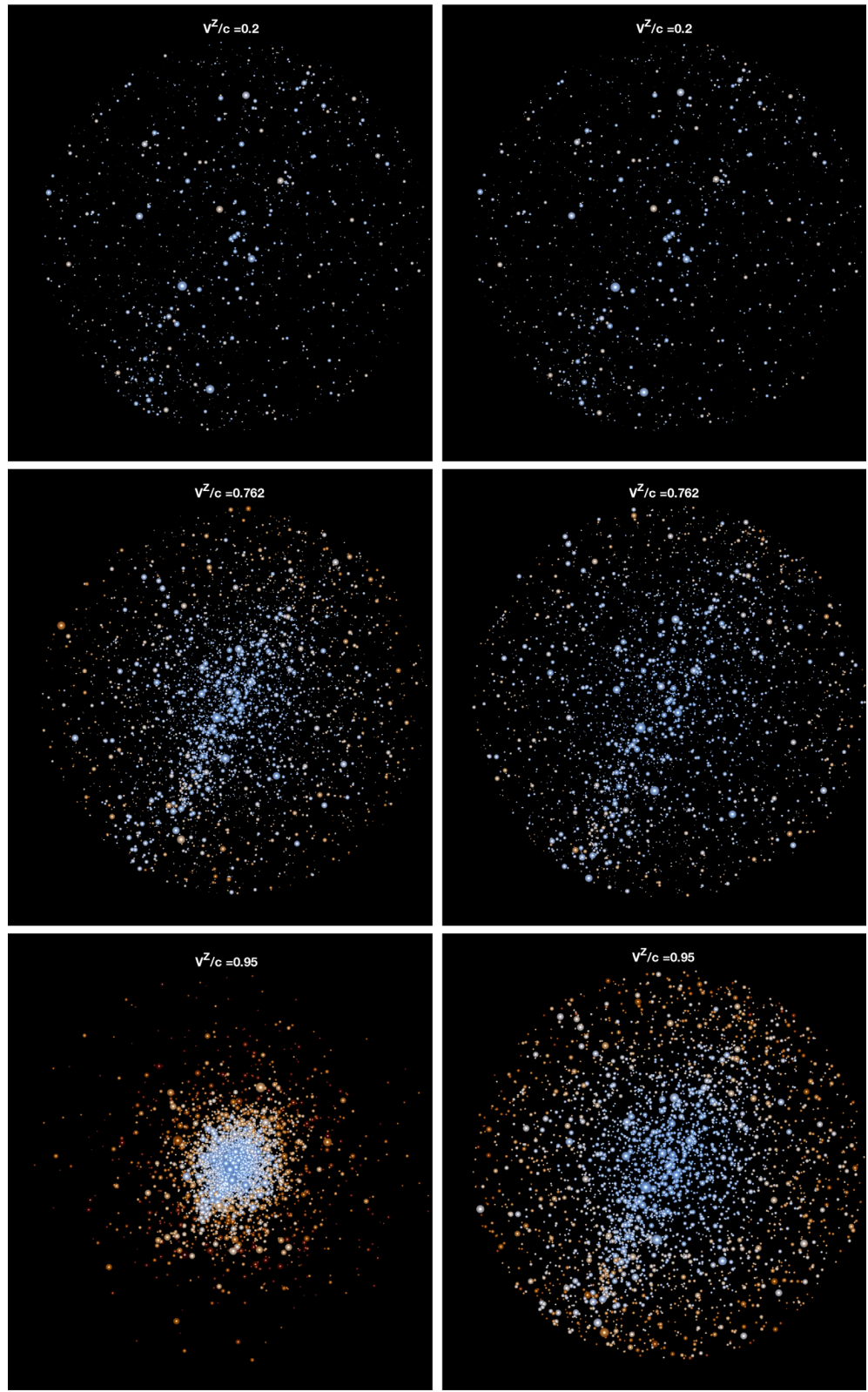

FIG. 10. Front views for travelers aboard photon rockets during their trip toward star Alnilam, at the center of the field of view, for increasing velocities (left: emission rocket; right: light sail).

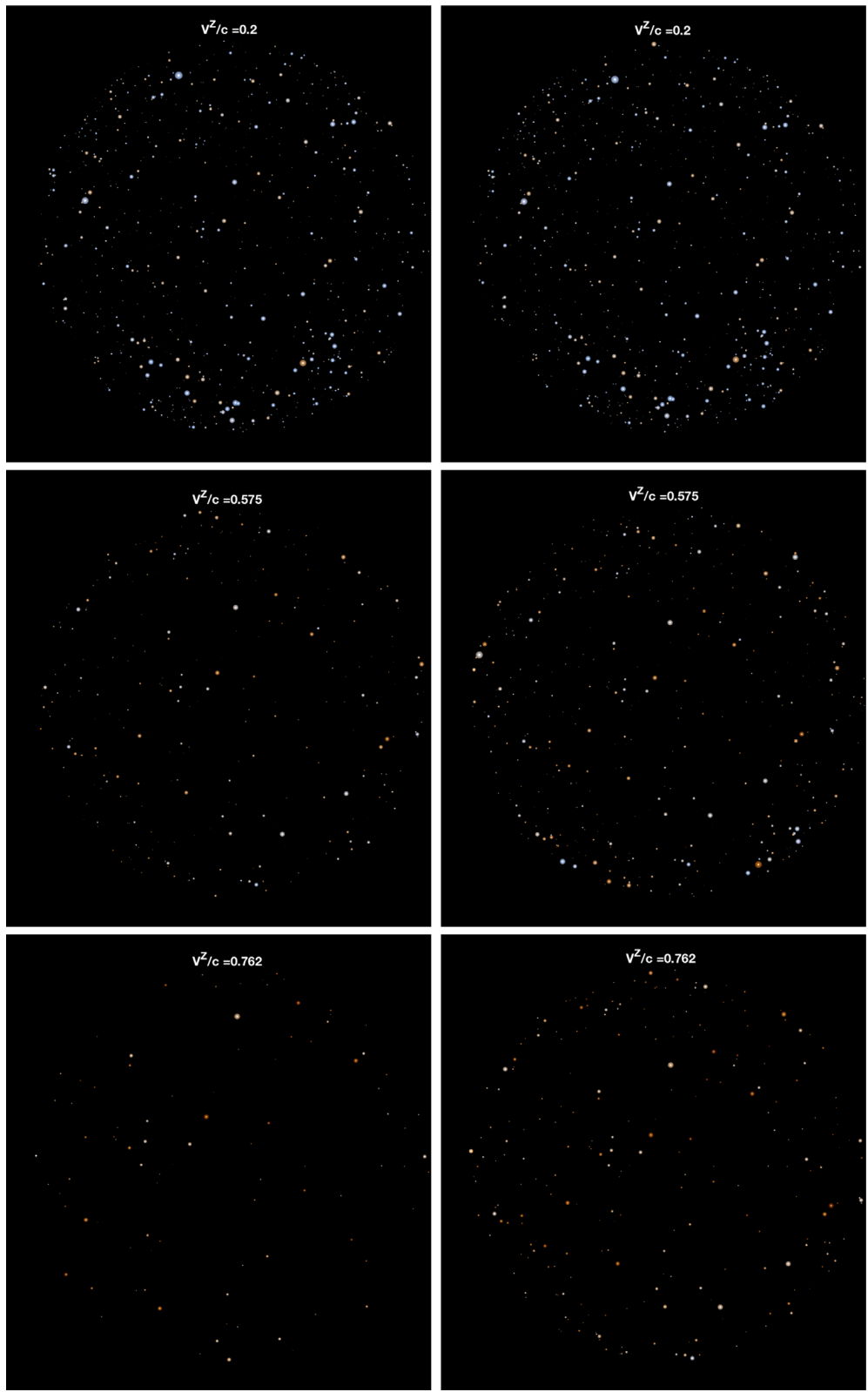

FIG. 11. Rear views for travelers aboard photon rockets during their trip toward star Alnilam as their velocity increases (left: emission rocket; right: light sail).

## IV. CONCLUSION

Interstellar travel at relativistic velocity ($V \lesssim c$) is not forbidden by any physical laws. Even better, it could be done in principle without the need of any massive propellant, without invoking any speculative physics and with trip duration significantly reduced by relativistic time dilation. Indeed, energy-momentum conservation in relativity allows propulsion using anisotropic emission or absorption of radiation, leading to accelerated trajectories on which time is slowed down as it is in the equivalent gravity field. These principles are actually well known but are often not correctly dealt with in discussions on this unfortunately somewhat controversial subject. Deep space propulsion by the reaction of radiation emission or absorption is at the basis of many plausible and working devices such as solar/laser-pushed sails, photonic propulsion, pure antimatter rockets, or radiative cooling rockets. Indeed, radiative rockets could even be propelled by the least noble type of energy, which is heat, through collimating the blackbody radiation of some hot radiator. It is the energy cost of interstellar travel that really prevents it from becoming a practical reality.

In addition, what has been missing so far is a rigorous physical modeling of radiation rockets in the framework of general relativity, which is unavoidable when one deals with accelerations, and this is the contribution of the present paper. Kinnersley's solution of general relativity gives a pointlike description of a photon rocket, although Einstein's equations reduce in this case to a single relation between the two functions of acceleration and mass in the metric and the (incoming and outgoing) radiation flux. To disentangle this problem, we use the energy-momentum conservation, which leads to the usual relativistic kinematics of the point particle, and derive specific models for light sails and radiative cooling rockets. We then applied these models to the practical example of interstellar trips to the Proxima Centauri star system, deriving important physical quantities for the acceleration, the variation of the rocket's inertial mass, and the time dilation aboard the rockets. It is shown how the strategy of ultralight laser sails is far more plausible than a manned radiative cooling rocket, notably from the point of view of the energy cost. Indeed, while the former would require a few days of operation of a single nuclear reactor, the latter would require about 15 times the annual world energy production…for a single mission.

Among the (numerous) technological challenges to achieve such an interstellar mission, there are the questions raised by telecommunication, course correction, navigation, and imaging at the destination. All these issues depend on how light rays are perceived by the traveler, and this depends on its past acceleration. By using the Hamiltonian formulation of geodesic flow, we have computed the trajectories of the incoming light rays for various types of radiation-powered rockets and derive the relativistic aberration (angular deviation of null geodesics) and Doppler effect (frequency shifts) experienced by the accelerated travelers. Our results extend the predictions of special relativity that are only valid for motion at constant velocity. It was also established analytically that, in the case of hyperbolic motion with constant norm of 4-acceleration, the aberration is strictly the same as in special relativity but not the Doppler effect. In general, different acceleration histories lead to stronger or weaker relativistic aberrations and Doppler shifts. We also built visualizations for the traveler's local celestial sphere that account for the modified aberration and Doppler effects found and showed what panoramas aboard an accelerated spaceship heading toward star Alnilam would look like.

The mass function of the Kinnersley metric was neglected while computing the modifications of relativistic beaming and the Doppler effect mentioned above. This is a rather safe assumption for the case of interstellar travels aboard radiation-powered rockets since we showed that the effects of acceleration largely dominate those of rocket inertial mass when the luminosity that powers the radiation rocket is much less than the huge value $c^5/G \approx 10^{52}$ W, sometimes referred to as Planck luminosity. Quite interestingly, the extreme amount of energy lost in gravitational radiation by binary black hole mergers would constitute an astrophysical application of Kinnersley metric where one could not neglect the effect of the mass function anymore. In particular, further studies should interestingly investigate the impact of the mass function on the Doppler effect and relativistic beaming associated to the radiation recoil of the merger. This can be done by using the cogeodesic equations derived here and applying them to mass and acceleration functions modeling the merger.

It is often (naively?) hoped that moving to other star systems will be our only escape if one day this planet becomes inhospitable. But actually, developing interstellar travel might well precipitate the exhaust of our planet resources. In our view, it is crucial that the difficulties and implications of interstellar travel are correctly taught, based on rigorous scientific argumentation. In addition, it seems to us that the time has come for starting the development of a technology demonstrator for a high-velocity radiation-powered rocket in the Solar System. The results of this paper are of direct application for the computation of the trajectory, the input-ouput transmissions to the probes, the relativistic aberration effects of image capture during a flyby, and course corrections of such a high-velocity demonstrator.

We can now pave the way for interstellar exploration with radiation-powered rockets, beyond the engineering sketches and theoretical exploratory works done so far. Hopefully, we will at last leave ourselves to this intimate experience common to all those who go out stargazing: the appeal of the stars.

ANDRÉ FÜZFA PHYS. REV. D **99,** 104081 (2019)

## ACKNOWLEDGMENTS

The author warmly thanks the anonymous referee and R. Lehoucq for their pertinent and encouraging remarks that helped improve the paper and its readability. This research was made during a scientific sabbatical funded by the Fonds National de la Recherche Scientifique F.R.S.-FNRS. This research used resources of the "Plateforme Technologique de Calcul Intensif (PTCI)" (http://www.ptci.unamur.be) located at the University of Namur, Belgium, which is supported by the FNRS-FRFC, the Walloon Region, and the University of Namur (Conventions No. 2.5020.11, GEQ U.G006.15, 1610468 et RW/GEQ2016). The PTCI is member of the "Consortium des Équipements de Calcul Intensif (CÉCI)" (http://www.ceci-hpc.be).

[1] A. Camus, *The Myth of Sisyphus*, translated from French by J. O'Brien, (Vintage Books, New York, 2018), http://www.vintagebooks.com/.

[2] Actually, the fastest object is the Voyager 1 probe with a record interplanetary cruise velocity of ∼17 km/s. Please note that other spacecrafts, like Parker Solar Probe, have reached a 10 times higher peak speed on some parts of their trajectory.

[3] K. F. Long, *Deep Space Propulsion–A Roadmap to Interstellar Flight* (Springer, New York, 2012).

[4] https://breakthroughinitiatives.org/initiative/3.

[5] E. Gourgoulhon, *Relativité Restreinte–Des particules à l'astrophysique* (EDP Sciences & CNRS Editions, Paris, France, 2010).

[6] W. Kinnersley, Field of an arbitrarily accelerating point mass, Phys. Rev. **186**, 1335 (1969).

[7] P. Vaidya, The gravitational field of a radiating star, Indian Acad. Sci. A **33**, 264 (1951); R. W. Lindquist, R. A. Schwartz, and C. W. Misner, Vaidya's radiating schwarzschild metric, Phys. Rev. **137**, B1364 (1965).

[8] H. Stephani, D. Kramer, M. MacCallum, C. Hoenselaers, and E. Herlt, *Exact Solutions of Einstein's Field Equations*, 2nd ed. (Cambridge University Press, Cambridge, England, 2003).

[9] W. B. Bonnor, The photon rocket, Classical Quantum Gravity **11**, 2007 (1994).

[10] In addition, it was shown in Ref. [13] that a radiation rocket can be powered by gravitational waves as well.

[11] W. B. Bonnor, Another photon rocket, Classical Quantum Gravity **13**, 277 (1996).

[12] J. Podolsky, Photon rockets in the (anti-)de Sitter universe, Phys. Rev. D **78**, 044029 (2008).

[13] W. B. Bonnor and M. S. Piper, The gravitational wave rocket, Classical Quantum Gravity **14**, 2895 (1997).

[14] J. Podolsky, Photon rockets moving arbitrarily in any dimension, Int. J. Mod. Phys. D **20**, 335 (2011).

[15] T. Damour, Photon rockets and gravitational radiation, Classical Quantum Gravity **12**, 725 (1995).

[16] S. Carlip, Aberration and the speed of gravity, Phys. Lett. A **267**, 81 (2000).

[17] C. Rubbia, Nuclear space propulsion with a pure electromagnetic thrust, CERN Report No. CERN-SL-2002-006, 2002, http://cds.cern.ch/record/541446.

[18] J. Ackeret, Zur Theorie der Raketen, Helv. Phys. Acta **19**, 103 (1946).

[19] It must be recalled here that the 4-acceleration is defined as the variation of the tangent vector with proper time and is not directly the conventional 3-acceleration with respect to some inertial observer, which is a related yet different quantity.

[20] D. Mihalas and B. Weibel-Mihalas, *Foundations of Radiation Hydrodynamics*, 2nd ed. (Dover, New York, 1999).

[21] C. W. Misner, J. A. Wheeler, and K. S. Thorne, *Gravitation* (Freeman, San Francisco, 1973).

[22] In addition, it was shown by Bonnor in Ref. [9] that quadratic curvature invariants depend only on the mass function $M$ and not on the acceleration function $\alpha$.

[23] P. B. Abbott *et al.* (LIGO Scientific and Virgo Collaborations), Observation of Gravitational Waves from a Binary Black Hole Merger, Phys. Rev. Lett. **116**, 061102 (2016).

[24] B. J. S. Dong *et al.*, ASASSN-15lh: A highly superluminous supernova, Science **351**, 257 (2016).

[25] J. L. Greenstein and M. Schmidt, The Quasi-Stellar Radio Sources 3C 48 and 3C 273, Astrophys. J. **140**, 1 (1964).

[26] C. J. Hogan, Energy flow in the Universe, NATO Sci. Ser. C **565**, 283 (2001).

[27] The local azimutal coordinate $\Phi$ can be identified to the coordinate $\varphi$ due to axial symmetry.

[28] A. Einstein, Zur Elektrodynamik bewegter Körper, Ann. Phys. (Berlin) **322**, 891 (1905).

[29] D. Hoffleit, *Catalogue of Bright Stars*, edited by D. Hoffleit, 3rd revised ed. (Yale University Observatory, New Haven, CT, 1964); data from fifth edition (1991) are available at http://tdc-www.harvard.edu/catalogs/bsc5.html.

[30] F. J. Ballesteros, New insights into black bodies, Europhys. Lett. **97**, 34008 (2012).

[31] http://cvrl.ioo.ucl.ac.uk/.